\documentclass[aps,prb,superscriptaddress,twocolumn,showpacs,floatfix]{revtex4-1}

\usepackage{graphicx}% Include figure files
\usepackage{dcolumn}% Align table columns on decimal point
\usepackage{bm}% bold math
\usepackage{color}
\usepackage[colorlinks=true,urlcolor=blue,citecolor=blue, linkcolor = blue]{hyperref}
\usepackage{amssymb,ulem,amsmath}
\def\be{\begin{equation}}
\def\ee{\end{equation}}
\def\ba{\begin{array}{lll}}
\def\ea{\end{array}}
\def\ber{\begin{eqnarray}}
\def\eer{\end{eqnarray}}
\newcommand{\bra}[1]{\left\langle #1\right|}
\newcommand{\ket}[1]{\left|#1\right\rangle}
\newcommand{\ketb}[1]{|#1\rangle}
\newcommand{\brab}[1]{\langle #1|}
\newcommand{\braket}[2]{\left\langle #1 | #2 \right\rangle}
\newcommand{\expect}[3]{\left\langle #1 \right| #2 \left| #3 \right\rangle}
\newcommand{\qql}{\textquotedblleft}
\newcommand{\qqr}{\textquotedblright}
\begin{document}
\title{Improving the Gutzwiller Ansatz with Matrix Product States}

\author{Sebastiano Peotta}
\author{Massimiliano Di Ventra}
\affiliation{Department of Physics, University of California-San Diego, La Jolla, CA 92093, USA}
\begin{abstract}
The Gutzwiller variational wavefunction (GVW) is commonly employed to capture correlation effects in condensed matter systems such as ferromagnets, ultracold bosonic gases, correlated superconductors, etc.
By noticing that the grand-canonical and number-conserving Gutzwiller Ans\"atze are in fact the zero-order approximation of an expansion in the truncation parameter $\epsilon$ of a Matrix Product State (MPS), we argue that MPSs, and the algorithms used to operate on them, are not only flexible computational tools but also a unifying theoretical framework that can be used to generalize and improve on the GVW. In fact, we show that a number-conserving GVW is less efficient in capturing the ground state of a quantum system than a more general MPS which can be optimized with comparable computational resources. Moreover, we suggest a corrected time-dependent density matrix renormalization group algorithm that ensures the conservation of the expectation value of the number of particles when a GVW or a MPS are not explicitly number-conserving. The GVW dynamics obtained with our algorithm compares very well with the exact one in 1D. Most importantly, the algorithm works in any dimension for a GVW. We thus expect it to be of great value in the study of the dynamics of correlated quantum systems.
\end{abstract}
%
%\pacs{67.10.Jn, 67.85.Lm}
%
\maketitle

\section{Introduction}
The Gutzwiller variational wavefunction\cite{Gutzwiller:1963,Bunemann:1998} (GVW)  has been an important tool in the analysis of various correlated quantum systems, ranging from ferromagnets
\cite{Hubbard:1963,Kanamori:1963,Borghi:2013} to ultracold bosonic gases\cite{Jaksch:1998}, to superconductors\cite{Anderson:1987, Zhang:1988, Edegger:2007, Bernevig:2003} (the so-called \qql Gutzwiller-correlated BCS wave function\qqr). In fact, in some cases -- such as transition metals --, its success in determining the dispersion of quasi-particle excitations has been confirmed experimentally by angle-resolved photoemission experiments\cite{Bunemann:2003} and de Haas-van Alphen data.\cite{Bunemann:2008}

More recently, the concept of Matrix Product States\cite{Ostlund:1995} (MPSs) has shedded new light on the Density Matrix Renormalization Group (DMRG) algorithm developed by White,\cite{White:1992,White:1993} which is very successful in describing the ground state properties of strongly interacting local Hamiltonians in one dimension. Indeed, \qql if we do quantum mechanics in the restricted state class provided by matrix product states, DMRG and other methods almost force themselves on us\qqr.\cite{tdmrg}
MPSs are not merely a convenient reformulation of White algorithm. In fact, they turn out to be a theoretical framework that allows for extensions and generalizations that would be too cumbersome to formulate within the old language of DMRG. For instance, DMRG suffers from severe limitations when extended to dimensions larger than one,\cite{tdmrg,Stoudenmire:2012} although
attempts in this direction have been explored with the so-called Projected Entangled Pair States (PEPS),\cite{Verstraete:2004} a simple generalization of MPSs.\cite{Nishino:2001}

%An approach that may provide a lot of physical information, tested against the exact solution in 1D but which is easily extendable to any dimension
%would be thus of great value in the study of correlated quantum systems.

In this paper, we first note that the grand-canonical and number-conserving Gutzwiller Ans\"atze are in fact the zero-order approximation of an expansion in the truncation parameter of a MPS. We use this simple but nontrivial observation in two different ways. First, we show that at essentially the same computational cost, a number-conserving GVW is less efficient in capturing the ground state of a quantum system than a more general MPS with comparable \textit{link dimension}, \textit{i.e.} with essentially the same number of variational parameters. Second, we propose a corrected Time-Dependent Density Matrix Renormalization Group algorithm\cite{tdmrg,White:2004,Vidal:2004,Daley:2004} (TDMRG) that conserves exactly the expectation value of the number of particles when a GVW or a MPS are not explicitly number-conserving.

Whereas the time-dependent Gutzwiller Ansatz has been used several times in the literature,\cite{Jaksch:2002,Snoek:2007,Wernsdorfer:2010,Natu:2011,Bernier:2012} to our knowledge  no explicit algorithm that allows for the \textit{exact} conservation of the number of particles has ever been described, and this prevents the comparison between the GVW dynamics and (quasi)-exact TDMRG simulations, which we can now provide in this work. Moreover, our algorithm works in {\it any dimension} for the GVW, and in one dimension for general MPSs that do not explicitly conserve the particle number or other conserved quantities corresponding to Abelian symmetries of the Hamiltonian.
Our algorithm can find applications in the study of correlated effects in quantum systems in dimension one and higher, both as a simple starting point and as a yard stick for more refined
calculations. Indeed, the GVW is a variational mean-field wavefunction, and it is expected to work even better with increasing dimension.

The paper is organized as follows. In Sec.~\ref{sec1} we recall the definition and properties of Matrix Product States, and how the use of symmetries greatly simplifies their computation. In Sec.~\ref{sec2} we show explicitly that the GVW is nothing other than a zero-order MPS.
In addition, we answer in the affirmative the question of whether it exists a MPS with a smaller link dimension that, at the same computational cost, better captures the ground state
wavefunction than a number-conserving GVW in 1D. In Sec.~\ref{sec3} we propose a novel number-conserving time-dependent Gutzwiller Ansatz, and show that it compares very well with the exact dynamics using TDMRG in 1D. The algorithm is not limited to the GVW but is in fact a corrected version of TDMRG that can be applied to general MPSs that do not explicitly conserve the number of particles. We provide a detail description of the algorithm only for the GVW since the corresponding more general version for MPSs is simply notationally more involved. Finally, we conclude in Sec.~\ref{conclusions}.

\section{Matrix product states}\label{sec1}
Although not limited to this case,  in the following we consider a quantum system defined on a open lattice of length $L$ where the lattice site $i$ can be occupied by a number $n_i = 0, \dots,+\infty$ of bosons. The Hamiltonian  is assumed to be the sum of nearest-neighbor terms which globally conserve the total number of particles $N = \sum_in_i$.
A typical example is the Bose-Hubbard Model\cite{Fisher:1989,Jaksch:1998,Zwerger:2003,Kunher:2000} (BHM)
\begin{equation}\label{eq:ebhm}
\mathcal{\hat H}_{\rm BHM} = -J\sum_i\left(\hat b_i^\dagger \hat b_{i+1}+\hat b_{i+1}^\dagger \hat b_i\right) + \frac{U}{2}\sum_i \hat n_i\,,
\end{equation}
containing a local interaction -- being a function of the number operator $\hat n_i= \hat b_{i}^\dagger \hat b_i$ -- that conserves the total number of particles. In the hopping term $\hat b_i^\dagger \hat b_{i+1}$ the operator $\hat b_{i+1}^\dagger$ destroys a particle on site $i+1$ and the operator $\hat b_i^\dagger$ creates one particle on site $i$, again conserving $\hat N$.

Before proceeding, let us first introduce some notations for the benefit of the reader. In the following a bold symbol $\bm{B}^{(i)}$ is a short hand for a tensor (or matrix) attached to site $i$ with two \textit{link indices}, denoted by  $\ell_{i-1}\,,\ell_{i}$ and components $B_{\ell_{i-1},\ell_{i}}$. The link index $\ell_i$ is relative to the link connecting site $i$ and site $i+1$ while $\ell_{i-1}$ refers to the link between site $i-1$ and site $i$.  Every  link index $\ell_{i}$  ranges  from  1 up to the \textit{link dimension} $m_i$, thus the matrix $\bm{B}^{(i)}$  has dimension $m_{i-1}\times m_i$.  $\bm{A}^{[n_i]}$ is a collection of tensors, one for each value of the occupation number $n_i$ attached to site $i$. For an inhomogeneous system $\bm{A}^{[n_i]}\neq \bm{A}^{[n_j]}$ in general, if $i\neq j$.
The dot \qql$\,\cdot\,$\qqr has the meaning of a contraction of a link index $\ell_i$
\begin{equation}
\left[\bm{A}^{[n_i]}\cdot \bm{A}^{[n_{i+1}]} \right]_{\ell_{i-1}\ell_{i+1}} = \sum_{\ell_{i} =1}^{m_i}
A^{[n_i]}_{\ell_{i-1}\ell_{i}}A^{[n_i]}_{\ell_{i}\ell_{i+1}}\,.
\end{equation}
A set $\{ \bm{A}^{[n_i]}\}_{i = 1,\dots,L}$ is called a Matrix Product State (MPS), an alternative way to specify a wavefunction $\ket{\Psi}$.  For  an arbitrary given set of occupancies  $\{n_i\}_{i = 1,\ldots,L}$ the complex number $\braket{\{n_i\}}{\Psi}$ is obtained by contracting all the link indices
\begin{equation}\label{eq:mps}
\begin{split}
&\braket{n_1,n_2,\ldots,n_{L-1},n_{L}}{\Psi} \\ &\qquad= \bm{A}^{[n_1]}\cdot \bm{A}^{[n_2]}\cdot \ldots\cdot \bm{A}^{[n_{L-1}]}\cdot \bm{A}^{[n_L]}\,.
\end{split}
\end{equation}
For open boundary conditions the leftmost link index $\ell_0$ and the rightmost one $\ell_L$ have dimensions $m_{0} = m_{L} = 1$ and do not need to be contracted.  Note that the exact ground state  wavefunction has always an {\it exact} MPS representation but with impractically large values of $m_i$ for long chains ($L\gtrsim 20$). Therefore, reducing the \textit{link dimension} $m_i$ with some sort of truncation procedure is the essential idea of MPS-based algorithms like DMRG.\cite{tdmrg}

It is computationally more convenient to restrict the Hilbert space only to the subspace of states with a fixed number of particles $N$ ($\braket{\{n_i\}}{\Psi} = 0$ for $\sum_in_i\neq N$).
 By explicitly enforcing this condition on the MPS (\ref{eq:mps}) results in a block structure for the matrices $\bm{A}^{[n_i]}$ described in the following. The $i$-th link is divided in symmetry multiplets labeled by an integer $\alpha_i$, which is the number of particles located on sites at the left of the $i$-th link. Each multiplet has a multiplicity $d_{\alpha_i}$, which can be zero, and the values of the link index $\ell_i$  can be grouped accordingly, namely  $\ell_i = (\alpha_i,k_{\alpha_i})$ with $1\leq k_{\alpha_i}\leq d_{\alpha_i}$ and $\sum_{\alpha_i}d_{\alpha_i} = m_i$.

We denote by $\bm{A}^{[n_i]}_{\alpha_{i-1}\alpha_{i}}$ the submatrix of $\bm{A}^{[n_i]}$ corresponding to the multiplets $\alpha_{i-1}$  on  the $(i-1)$-th link and $\alpha_i$ on the $i$-th link. The definition of $\alpha_i$ leads to the condition $\alpha_i = n_i + \alpha_{i-1}$ for $\bm{A}^{[n_i]}_{\alpha_{i-1}\alpha_{i}}$ to be
nonzero, thus large blocks of $\bm{A}^{[n_i]}$ are zero and the size of the MPS is greatly reduced. At the left boundary only the sector $\alpha_0 = 0$ has nonzero multiplicity $d_{\alpha_0 = 0}=1$ and the same holds for $\alpha_L = N$ at the right boundary.
A MPS with a block structure induced by a $U(1)$ (abelian) symmetry is easier to optimize since the dimension of the local eigenvalue problem to be solved is drastically reduced.\cite{tdmrg} Moreover one can perform several singular value decomposition (SVD) on each block instead of a more time consuming single SVD on a large matrix, an operation routinely performed during an imaginary- or real-time evolution.\cite{tdmrg}

\begin{figure*}
\begin{center}
\includegraphics[angle=0,width=16cm]{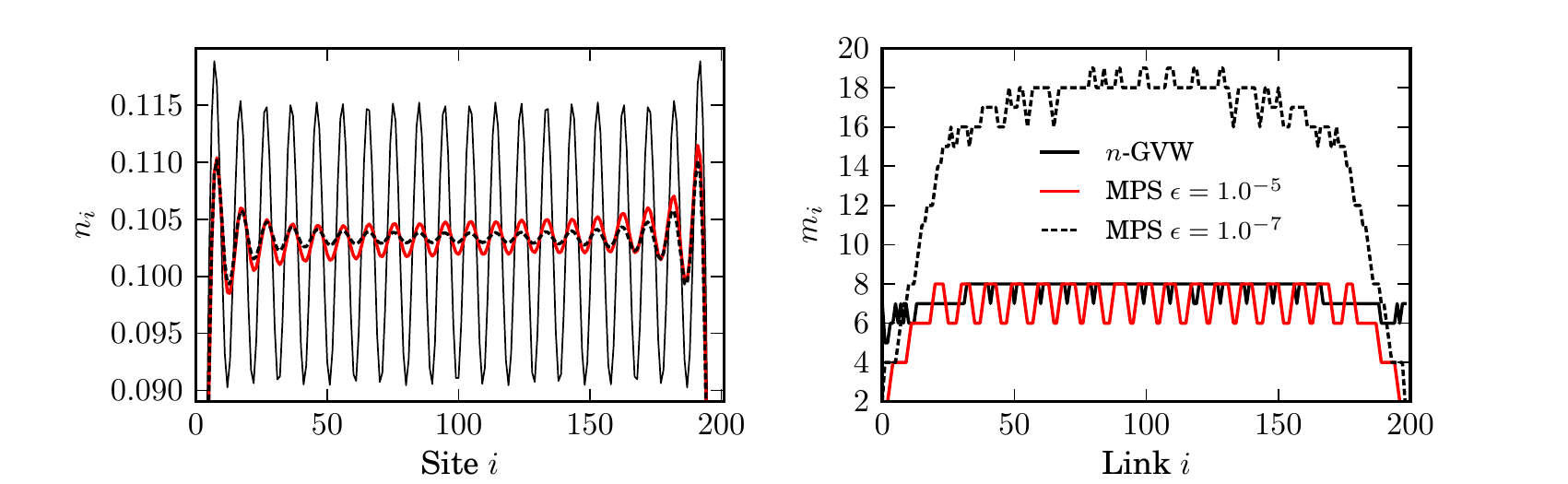}
\caption{\label{cap:fig1} (Color online) Left: density profiles obtained by means of a $n$-GVW and MPSs with different truncation parameter $\epsilon$. The $n$-GVW shows pronounced density oscillations with constant amplitude throughout the chain.  On the contrary the more general MPS Ansatz is able to capture the quantum fluctuations that lead to a suppression of the oscillations. Right: link dimension $m_i$ for the same variational wavefunctions used for the left panel. Note that the  $n-$GVW has a link dimension comparable to that of a MPS with $\epsilon = 10^{-5}$, but the latter is more efficient in describing the ground state as it can be see from the variational ground state energy shown in Fig.~\ref{cap:fig2}.}
\end{center}
\end{figure*}

\begin{figure}
\includegraphics{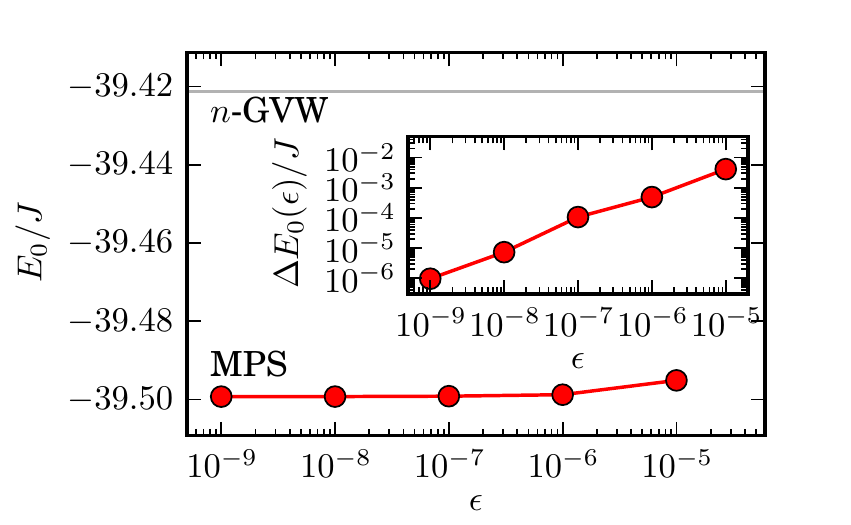}
\caption{\label{cap:fig2} (Color online) Ground-state energy $E_0/J$ given by the $n-$GVW  (grey line) and by MPSs with different discarded weights $\epsilon$ (red dots). The inset shows the error in the  ground-state energy $\Delta E_0(\epsilon) = (E_0(\epsilon)-E_0(10^{-12}))/J$ for MPSs with different discarded weight $\epsilon$. The MPS with $\epsilon = 10^{-12}$ is taken as a reference for the exact ground state.}
\end{figure}

\section{Gutzwiller Ansatz as a Matrix Product State}\label{sec2}
For a bosonic system a commonly employed approximation is the \textit{grand-canonical} Gutzwiller variational wavefunction ($\mu$-GVW)
\begin{equation}\label{eq:gran_GVW}
\ket{\Psi}_{\mu\rm-GVW} = \bigotimes_i\ket{\Psi_i} \qquad \ket{\Psi_i} = \sum_ic^{(i)}_{n}\ket{n_{i}}\,.
\end{equation}
The state $\ket{n_i}$ is an eigenstate of the number operator $\hat n_i\ket{n_i} = n_i\ket{n_i}$ and the $c_n^{(i)}$ are arbitrary variational parameters.

We immediately note that the grand-canonical Gutzwiller Ansatz is the most general MPS with link dimension $m_i = 1$ for every link $i$, since the matrix $\bm{A}^{[n_i]} = c_{n}^{(i)}$ becomes a simple scalar. From a $\mu$-GVW one can easily derive a \textit{canonical} (number conserving) Gutzwiller Ansatz\cite{Jaksch:2002} which we call ($n$-GVW)
\begin{equation}
\begin{split}
\ket{\Psi}_{n\rm-GVW} &=
\frac{\mathcal{P}_{N}\ket{\Psi}_{\mu\rm-GVW}}
{\lVert\mathcal{P}_{N}\ket{\Psi}_{\mu\rm-GVW}\rVert}\\
&\sim \int_0^{2\pi}d\phi\, e^{-iN\phi}\bigotimes_{i}\left(e^{i\hat n_i\phi}\ket{\Psi_i}\right)\,.
\end{split}
\end{equation}
$\mathcal{P}_N$ is the projector on the subspace with $N$ particles. The $n$-GVW can be expressed as a number-conserving MPS where all the nonzero blocks ($\alpha_i-\alpha_{i-1} = n_i$) are  scalars $\bm{A}_{\alpha_i-1,\alpha_i}^{[n_i]} = c_{n}^{(i)}$, \textit{i.e.}, all symmetry multiplets are in fact singlets ($d_{\alpha_i} = 1$ for all $i$ and $\alpha_i$). Note that not every MPS with identically $d_{\alpha_i} = 1$ can be obtained by projecting a $\mu$-GVW, since in the latter case the following constrains are enforced
\begin{equation}
\bm{A}^{[n_i]}_{\alpha_{i-1},\alpha_{i}} = \bm{A}^{[n_i]}_{\beta_{i-1},\beta_{i}} = c_n^{(i)}\qquad \alpha_i-\alpha_{i-1} = \beta_{i}-\beta_{i-1} = n_i\,.
\end{equation}

It is quite cumbersome to deal with such constrains numerically, thus in the following we \textit{redefine} a  canonical Gutzwiller Ansatz as the most general number-conserving MPS where the blocks $\bm{A}_{\alpha_{i-1},\alpha_i}^{[n_i]}$ are scalars ($d_{\alpha_i} ,d_{\alpha_{i-1}} \leq 1$). This class of states can be easily dealt with the usual methods of DMRG.\cite{tdmrg} An important observation is that particle number conservation alone has the nontrivial effect of increasing the amount of entanglement of the trial wavefunction. For a $\mu-$GWV $m_i = 1$ for all links and the block entropy\cite{tdmrg} is identically zero.
On the other hand for a $n-$GVW one has $m_i = \sum_{\alpha_i=0}^{N+1}d_{\alpha_i} \leq N+1$. Thus the block entropy is bounded by $\log(N+1)$ and nonzero in general, unless each particle is localized on a single site. This implies that a $n-$GVW is computationally more expensive than a $\mu-$GVW.

An interesting question is then if a more general MPS -- where the degeneracies $d_{\alpha_i}$ of the symmetry multiplets are not bounded ($d_{\alpha_i} \nleq 1$) --  can better capture the ground-state wavefunction at a {\it comparable} computational cost, where the computational cost of a MPS is roughly quantified by the link dimension $m_i$. In the following, we will indeed show that the answer is affirmative.

To show this explicitely, we consider $N = 20$ particles in a lattice of $L=200$ sites with Hamiltonian given by Eq.~(\ref{eq:ebhm}) with $U/J = 2.0$ and, by using the standard DMRG algorithm, we optimize the $n-$GVW. The results are compared with MPSs with fixed discarded weight\cite{tdmrg} $\epsilon$, meaning that after each SVD the discarded singular values satisfy
\begin{equation}
\sum_{\rm discarded}\sigma_i^2 < \epsilon\,.
\end{equation}

In Fig.~\ref{cap:fig1} we show the density profiles $n_i$ and the link dimensions $m_i$ relative to a $n-$GVW and MPSs.
Pronounced  density oscillations with constant amplitude are visible in the profile obtained with an optimized $n-$GVW. These oscillations are in fact a charge density wave induced by the repulsive Hubbard interaction $U\hat n_i(\hat n_i - 1)$. The more flexible MPSs are able to describe the quantum fluctuations occurring in the system, something which is beyond the capabilities of the Gutzwiller Ansatz, which is essentially a mean-field approximation. Indeed for the MPS with $\epsilon = 10^{-5}$ the charge density order is destroyed by quantum fluctuations, and the oscillations decay from the boundary towards the middle of the chain. Further decreasing $\epsilon$ to $10^{-7}$ leads to a further suppression of the oscillations. On the right the corresponding data for the link dimensions $m_i$ for the trial wavefunction shows that the $n-$GVW is in fact as computationally difficult to calculate as a MPS with a discarded weight as large as $10^{-5}$ (quite a poor value for current DMRG standards) and much better results can be obtained with the latter since it has a much lower energy (see Fig.~\ref{cap:fig2}).

\begin{figure}
\includegraphics{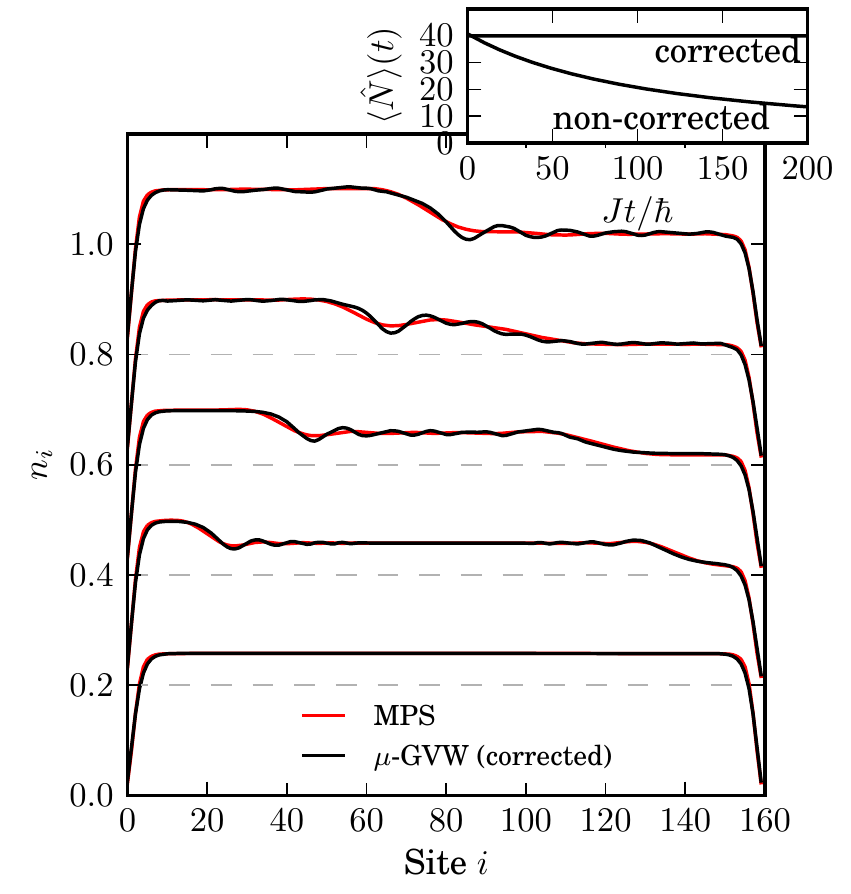}
\caption{\label{cap:fig3} (Color online) Time evolution of the density profile $n_i(t)$ at times $t = 0,\,30,\,60,\,90,\,120\, \hbar/J$ calculated after quenching the complex hopping $-J\sum_i\left(e^{i\phi(t)}\hat b_i^\dagger \hat b_{i+1}+e^{-i\phi(t)}\hat b_{i+1}^\dagger \hat b_i\right)$ from $\phi(t = 0) = 0$ to $\phi(t > 0) = 0.05$ in the Bose-Hubbard model (Eq.~\ref{eq:ebhm}) with $U/J=1.0$. The profiles have been shifted vertically for clarity. These data have been obtained both with a $\mu-$GVW (black line) employing the TDMRG algorithm corrected in order to preserve the expectation value of the particle number $\langle \hat N\rangle$ (see Sec.~\ref{sec3}) and with a MPS (red line)  with discarded weight $\epsilon = 10^{-10}$, which can be considered as numerically exact (see Ref.~\onlinecite{phase_induced_transport} for details). In the inset we compare $\langle \hat N \rangle (t)$ as function of time for a $\mu-$GVW evolved  with the standard TDMRG algorithm with and without correction. Without the correction $\langle \hat N\rangle(t)$ descreases rapidly.}
\end{figure}

\section{Number-conserving time-dependent Gutzwiller Ansatz}\label{sec3}
From the above we conclude that at least in one dimension a number-conserving Gutzwiller Ansatz is a rather poor choice, since a general MPS produces much better results. However number-conserving MPSs require a substantial amount of bookkeeping resulting in quite complex computer programs. For some applications one may instead consider a $\mu-$GVW which has the least possible computational cost. However, when a $\mu-$GVW is used to study the time evolution there is no guarantee that the expectation value of the particle number operator $\bra{\Psi(t)} \hat N\ket{\Psi(t)}_{\mu\rm-GVW}$  will remain constant in time. By observing that a $\mu-$GVW is nothing other than a particular instance of MPS, we then propose a modified TDMRG that {\it guarantees} that $\langle \hat N \rangle$ is \textit{exactly} constant in time.

For this purpose, let us first briefly review the TDMRG algorithm. If the Hamiltonian $\mathcal{\hat H}$ used to perform the time evolution can be broken into two pieces
 $\mathcal{\hat H} = \mathcal{\hat H}_{\rm even}+ \mathcal{\hat H}_{\rm odd}$, each a sum of mutually commuting operators $\hat h_{i,i+1}$ acting on two adjacent sites,  then the operator that evolves the system in time for a step $\Delta t$ can be approximated as (Trotter decomposition)
\begin{equation}
 \mathcal{U}(\Delta t)\approx e^{{i\Delta t\, \mathcal{ \hat H}_{\rm odd}}/2} e^{i\Delta t\, \mathcal{ \hat H}_{\rm even}} e^{{i\Delta t\, \mathcal{ \hat H}_{\rm odd}}/2}\,,
\end{equation}
and it is possible to separately apply
\begin{gather}
\exp({i\Delta t\, \mathcal{ \hat H}_{\rm even}}) = \bigotimes_{i}\exp(i\Delta t\,\hat h_{2i,2i+1})\,, \\
\exp({i\Delta t\, \mathcal{ \hat H}_{\rm odd}}) = \bigotimes_{i}\exp(i\Delta t\,\hat h_{2i-1,2i})\,,
\end{gather}
 by updating two MPS matrices at a time
\begin{equation}
\begin{split}
&\bm{M}^{[n'_i,n'_{i+1}]}= \\& \sum_{n_i,n_{i+1}}
 [\exp({i\Delta t\, \mathcal{ \hat H}_{\rm odd(even)}})]^{n'_i,n'_{i+1}}_{n_i,n_{i+1}}\bm{A}^{[n_i]}\cdot \bm{A}^{[n_{i+1}]}\,.
\end{split}
\end{equation}
In order to keep the MPS dimensions bounded one finds the best rank $m_i$ approximation $\bm{B}^{[n_i]}\cdot \bm{B}^{[n_{i+1}]}$ of $\bm{M}^{[n_i,n_{i+1}]}$ by minimizing the functional
\begin{equation}
\left\lVert \bm{M}^{[n_i,n_{i+1}]} - \bm{B}^{[n_i]}\cdot \bm{B}^{[n_{i+1}]}\right\rVert^2\,,
\end{equation}
where $\bm{B}^{[n_i]}\;(\bm{B}^{[n_{i+1}]})$ are matrices with dimension $m_{i-1}\times m_{i}\;(m_{i}\times m_{i+1})$.
The optimal solution is then obtained by retaining the largest $m_i$  singular values of $\bm{M}^{[n_i,n_{i+1}]}$ (for more details see Ref.~\onlinecite{tdmrg}).
The expectation value of $\hat N$ calculated with the new MPS $\bm{A}^{[n_i]}\cdot \bm{A}^{[n_{i+1}]} \to \bm{M}^{[n_i,n_{i+1}]}$ is unchanged since $\exp({i\Delta t\, \mathcal{ \hat H}_{\rm odd(even)}})$ is a number-conserving operator but this is not necessarily true for the low rank approximation $\bm{A}^{[n_i]}\cdot \bm{A}^{[n_{i+1}]} \to \bm{B}^{[n_i]}\cdot \bm{B}^{[n_{i+1}]}$.

In the following we specialize to the case of a $\mu-$GVW but nothing  prevents to extend the algorithm presented in the following to a MPS that does not explicitly conserves the number of particle. A working implementation of the algorithm for the GVW is provided in Ref.~\onlinecite{code}.

We propose to minimize the functional
\begin{equation}\label{eq:func}
\begin{split}
&\left\lVert \mathcal{\hat U}\ket{\Psi_i}\ket{\Psi_{i+1}} - \ket{\Phi_{i}}\ketb{\Phi_{i+1}}\right\rVert^2 \\&- \mu \left(\expect{\Phi_i}{\hat n_i}{\Phi_i} + \expect{\Phi_{i+1}}{\hat n_{i+1}}{\Phi_{i+1}}\right)
\end{split}
\end{equation}
with respect to $\ket{\Phi_{i}}$ and $\ketb{\Phi_{i+1}}$ which are assumed to be normalized (the normalization condition can be enforced by additional Lagrange multipliers that will be introduced below). The operator $\mathcal{\hat U} = \bm{1} + O(\Delta t)$ is a generic evolution operator acting on two sites and sufficiently close to the identity. The Lagrange multiplier $\mu$ is introduced in order to enforce the condition of particle number conservation
\begin{equation}\label{eq:conserv}
\begin{split}
&\expect{\Phi_i}{\hat n_i}{\Phi_i} + \expect{\Phi_{i+1}}{\hat n_{i+1}}{\Phi_{i+1}} \\&= \expect{\Psi_i}{\hat n_i}{\Psi_i} + \expect{\Psi_{i+1}}{\hat n_{i+1}}{\Psi_{i+1}} \,.
\end{split}
\end{equation}

Varying the functional~(\ref{eq:func}) with respect to $\bra{\Phi_i}$ and $\brab{\Phi_{i+1}}$ gives two coupled equations
\begin{gather}\label{eq:1}
\ketb{\Phi_i} = \frac{\varepsilon_1}{1+\lambda \hat n_i}\brab{\Phi_{i+1}} \mathcal{\hat U}\ketb{\Psi_i}\ketb{\Psi_{i+1}}\,, \\
\label{eq:2}
\ketb{\Phi_{i+1}} = \frac{\varepsilon_2}{1+\lambda \hat n_{i+1}}\bra{\Phi_i} \mathcal{\hat U}\ketb{\Psi_i}\ketb{\Psi_{i+1}}\,.
\end{gather}
The additional Lagrange multipliers $\varepsilon_1$ and $\varepsilon_2$ are used to preserve the normalization condition
\begin{equation}\label{eq:norm}
\braket{\Phi_i}{\Phi_i} = \braket{\Phi_{i+1}}{\Phi_{i+1}} = 1\,.
\end{equation}
 The parameter $\lambda$ is proportional to $\mu$ and must be adjusted to ensure the validity of Eq.~(\ref{eq:conserv}). The coupled equations~(\ref{eq:conserv}), (\ref{eq:1}), (\ref{eq:2}) and (\ref{eq:norm}) in the unknowns $\ketb{\Phi_i}\,,\ket{\Phi_{i+1}},\,\lambda,\,\varepsilon_1,\,\varepsilon_2$ can be solved iteratively. The first step in the iterative procedure is to first solve the equations for $\lambda = 0$. This is nothing else than the usual TDMRG algorithm where $\hat U\ket{\Psi_i}\ket{\Psi_{i+1}}\sim \ketb{\Phi^{(0)}_i}\ketb{\Phi^{(0)}_{i+1}}$ is approximated by truncating to the largest $m_i$ singular values ($m_i = 1$ in the case of a $\mu-$GVW). The couple of states obtained in such a way are the first of a sequence $\ketb{\Phi^{(j)}_i}\ketb{\Phi^{(j)}_{i+1}}$ constructed as follows.

Define the non-normalized states
\begin{gather}
\ket{\phi_i(\lambda)} = \frac{1}{1+\lambda \hat n_i}\brab{\Phi^{(j)}_{i+1}}\mathcal{\hat U}\ketb{\Psi_i}\ketb{\Psi_{i+1}}\,, \\
\ket{\phi_{i+1}(\lambda)}  = \frac{1}{1+\lambda \hat n_{i+1}}\brab{\Phi^{(j)}_i}\mathcal{\hat U}\ketb{\Psi_i}\ketb{\Psi_{i+1}}\,,
\end{gather}
and find the solution $\lambda^*$ of the equation
\begin{equation}\label{eq:lambda}
\begin{split}
f(\lambda) &\equiv \frac{\expect{\phi_i(\lambda)}{\hat n_i}{\phi_i(\lambda)}}{\left\lVert \ketb{\phi_i(\lambda)}\right\rVert^2}+\frac{\expect{\phi_{i+1}(\lambda)}{\hat n_{i+1}}{\phi_{i+1}(\lambda)}}{\left\lVert \ketb{\phi_{i+1}(\lambda)}\right\rVert^2}\\ &= \expect{\Psi_i}{\hat n_i}{\Psi_i} + \expect{\Psi_{i+1}}{\hat n_{i+1}}{\Psi_{i+1}}\,.
\end{split}
\end{equation}
Thus the normalized states
\begin{gather}
\ketb{\Phi^{(j+1)}_{i}} = \frac{\ketb{\phi_i(\lambda^*)}}{\lVert\ketb{\phi_i(\lambda^*)}\rVert}\,,\qquad
\ketb{\Phi^{(j+1)}_{i+1}} =  \frac{\ketb{\phi_{i+1}(\lambda^*}}{\lVert\ketb{\phi_{i+1}(\lambda^*)}\rVert}\,,
\end{gather}
preserve $\langle \hat n_i+\hat n_{i+1}\rangle$ (Eq.~\ref{eq:conserv}) and are  a rank-1 approximation of the evolved two-site state $\mathcal{\hat U}\ket{\Psi_i}\ket{\Psi_{i+1}} $.

We cannot prove in general that the sequence of states just defined converges to the solution of Eqs.~(\ref{eq:conserv}), (\ref{eq:1}), (\ref{eq:2}) and (\ref{eq:norm}), but we have observed that this is always the case when $\mathcal{\hat U}$ is a unitary operator close to the identity as in TDMRG simulations. In this case, one can linearize $f(\lambda)$ in $\lambda = 0$ and obtain a very good guess for the solution of Eq.~(\ref{eq:lambda}) ($\Delta \hat n_i = \hat n_i-\langle \hat n_i\rangle$)
\begin{widetext}
\begin{equation}
\lambda_{\rm guess} = -\frac{1}{2}\frac{ \expect{\phi_i(0)}{\hat n_i}{\phi_i(0)}
+\expect{\phi_{i+1}(0)}{\hat n_{i+1}}{\phi_{i+1}(0)} - \expect{\Psi_i}{\hat n_i}{\Psi_i} - \expect{\Psi_{i+1}}{\hat n_{i+1}}{\Psi_{i+1}}}
{\expect{\phi_i(0)}{(\Delta \hat n_i)^2}{\phi_i(0)}+\expect{\phi_{i+1}(0)}{(\Delta \hat n_{i+1})^2}{\phi_{i+1}(0)}}\,.
\end{equation}
\end{widetext}

In our simulations we always found that $\lambda^*\in[0,2\lambda_{\rm guess}]$ and that convergence is achieved in $\lesssim 5$ steps.
In Fig.~\ref{cap:fig3} we show a test of the algorithm just presented against a much more numerically demanding simulation performed using a number-conserving MPS. The density profile $n_i$ is shown after quenching a complex hopping term $-J\sum_i\left(e^{i\phi(t)}\hat b_i^\dagger \hat b_{i+1}+e^{-i\phi(t)}\hat b_{i+1}^\dagger \hat b_i\right)$ from $\phi(t = 0) = 0$ to $\phi(t > 0) = 0.05$ in the Hamiltonian~(\ref{eq:ebhm}) with $U/J = 1.0$. This amounts to a finite momentum delivered to the system. This kind of quench has been studied in Ref.~\onlinecite{Chien:2013,phase_induced_transport}.

The $\mu-$GVW evolved with the corrected TDMRG algorithm is able to capture the main features of the 1D dynamics which is quite interesting given the drastic approximation. Clearly, the MPS result shows less pronounced density oscillations since quantum fluctuations are captured by the variational wavefunction contrary to the $\mu-$GVW, as discussed in Sec.~\ref{sec3} and in the caption of Fig.~\ref{cap:fig1}. The inset of Fig.~\ref{cap:fig3} shows that without the correction the simulations are not reliable since $\langle \hat N\rangle$ decreases rapidly in time producing wrong results.

\section{Conclusions}\label{conclusions}
In summary, we have shown that the common grand-canonical and number-conserving Gutzwiller Ans\"atze are simply the zero-order approximation of an expansion in the truncation parameter $\epsilon$ of a Matrix Product State (MPS). This is an alternative point of view with respect to Ref.~\onlinecite{Kotliar:1986}  where the GVW can be derived as a saddle point approximation of an appropriate functional integral. Moreover, although equally efficient from a computational point of view, we have explicitly shown that a number-conserving GVW is less efficient in capturing the ground state of a quantum system than a more general MPS.
We believe that this is an important point to make since, despite the crudeness of the approximation, the Gutzwiller wavefunction is still a workhorse for the study of correlation effects in quantum systems, and it may be possible that even in higher dimensions a more general MPS (or PEPS\cite{Verstraete:2004} in this case) of a relatively small and manageable size provides better results.

On the other hand, since the GVW can be easily applied to correlated quantum systems in higher dimensions -- and indeed the approximation improves
with increasing dimension --, we have suggested a novel time-evolution algorithm to {\it exactly} conserve the expectation value of the number of particles when a GVW or a MPS are not explicitly number-conserving. This algorithm can find application in one dimension for MPSs, and in dimensions higher than one for the GVW. Most importantly, we have found that the GVW dynamics obtained with our algorithm compares very well with the exact one in 1D. As subsequent projects it would then be of great interest to apply our algorithm to correlated quantum systems in higher dimensions and compare with experiments or other theoretical methods such as DMFT or LDA+U approximation schemes.

\begin{acknowledgments}
This work has been supported by DOE under Grant No. DE-FG02-05ER46204. The Gutzwiller code developed for this work is made available at Ref.~\onlinecite{code}. The TDMRG code based on MPS has been developed in collaboration with Davide Rossini at the Scuola Normale Superiore, Pisa (Italy).
\end{acknowledgments}

\end{document}